\documentclass[%
 reprint,
 amsmath,amssymb,
 aps,
 pra,
]{revtex4-1}

\usepackage{graphicx}
\usepackage[varg]{txfonts}   
\usepackage{enumerate}
\usepackage{setspace}

\def\crta{\vrule height1.41ex depth-1.27ex width0.34em}
\def\dj{d\kern-0.36em\crta}
\def\Crta{\vrule height1ex depth-0.86ex width0.4em}
\def\Dj{D\kern-0.73em\Crta\kern0.33em}
\def\crta{\vrule height1.41ex depth-1.27ex width0.34em}
\def\dj{d\kern-0.36em\crta}
\def\Crta{\vrule height1ex depth-0.86ex width0.4em}
\def\Dj{D\kern-0.73em\Crta\kern0.33em}
\begin{document}

\title{Vector generation of contextual sets}

\author{Mladen Pavi{\v c}i{\'c}}
\email{mpavicic@irb.hr}
\homepage{http://www.irb.hr/users/mpavicic}
\affiliation{Department of Physics---Nanooptics, Faculty of Math. and Natural Sci.~I, Humboldt University of Berlin, Germany}
\affiliation{Center of Excellence CEMS, Photonics and Quantum Optics
  Unit, Ru\dj er Bo\v skovi\'c Institute, Zagreb, Croatia.}
\author{Norman D. Megill}
\affiliation{Boston Information Group, Lexington, MA 02420, U.S.A.}

\date{May 6, 2019}


\begin{abstract}
As quantum contextuality proves to be a necessary resource for
universal quantum computation, we present a general method
for vector generation of Kochen-Specker (KS) contextual sets in
the form of hypergraphs. The method supersedes all three previous
methods: (i) fortuitous discoveries of smallest KS sets,
(ii) exhaustive upward hypergraph-generation of sets,
and (iii) random downward generation of sets from fortuitously
obtained big master sets. In contrast to previous
works, we can generate master sets which contain all possible KS
sets starting with nothing but a few simple vector components.
From them we can readily generate all KS sets obtained in the last
half a century and any specified new KS sets. Herewith we can
generate sufficiently large sets as well as sets with definite
required features and structures to enable varieties of different
implementations in quantum computation and communication.
\end{abstract}

\maketitle

\section{\label{sec:intro}Introduction}

It has recently been recognized that ``contextuality [can
serve] as a [quantum] computational resource'' \cite{magic-14}.
In particular, in \cite{magic-14} it is shown ``that
what makes magic states [superposed states of qubits
initialized by quantum stabilizers] special is precisely their
contextuality. Specifically, they find that magic states possess
exactly the properties needed to prove that quantum physics is
contextual using an experimental test that relies only on
stabilizer operations. That is, the authors demonstrate that
this particular measurable aspect of quantum
weirdness---contextuality---is the source of a quantum computer's
power'' \cite{bartlett-nature-14}.

Further elaboration and future implementation of contextual sets
within such a framework of providing a computational resource
for quantum computation would require an optimal way for their
massive generation with desired properties and structures. In this
paper we give a method for generating the kind of contextual sets that
so far has been explored the most---the so-called Kochen-Specker
(KS) sets, which are constructive proofs of the KS
theorem \cite{koch-speck} which ``implies the impossibility
of explaining the statistical predictions of quantum theory in a
natural way. In particular, the actual outcome observed under a
quantum measurement cannot be understood as simply revealing a
preexisting value[s]'' \cite{magic-14} of a classical theory or a
classical binary, 0-1, calculation.  

KS sets have been implemented in a series of
recent experiments. Four dimensional (4D) KS sets have been
carried out recently, using photons
\cite{simon-zeil00,michler-zeil-00,amselem-cabello-09,liu-09,d-ambrosio-cabello-13,ks-exp-03}, neutrons
\cite{h-rauch06,cabello-fillip-rauch-08,b-rauch-09},
trapped ions \cite{k-cabello-blatt-09}, and
solid state molecular nuclear spins  \cite{moussa-09}.
Experiments have also been done with 6D systems via six
paths \cite{lisonek-14,canas-cabello-14} and with photons in an
8D space \cite{canas-cabello-8d-14}.

In quantum communication and quantum cryptography, KS sets 
are used for protecting \cite{cabello-dambrosio-11} and securing
\cite{nagata-05} quantum key distribution (QKD) protocols.
Quantum contextuality can be used to reveal quantum
nonlocality \cite{cabello-10}.
KS sets can serve as generators of higher-order generalized
orthoarguesian lattices in lattice theory
\cite{bdm-ndm-mp-fresl-jmp-10,mp-7oa}.
Also, operator-generated 4D complex KS sets provide a design for
constructing quantum gates \cite{waeg-aravind-jpa-11}.

The features of the previous approaches to KS sets in the
literature are as follows. 

The smallest KS sets
\cite{peres,kern,kern-peres,cabell-est-96a,cabell-garc98,b-rauch-09,cabello-pla09,d-ambrosio-cabello-13,lisonek-14,planat-12,planat-saniga-12,waeg-aravind-pra-17} are of just historical relevance because
practically all of them in even dimensional (from 4D up to 32D)
Hilbert spaces are already known and because all of them are
by-products of more comprehensive KS set generation. On the
other hand, experimentally, the complexity of implementation
grows only linearly with the complexity of the sets and it has
been found out that the simplest of KS sets often do not possess
features that bigger KS sets exhibit, like, e.g., the so-called
$\delta$-feature, absence of real coordinatization, absence of
parity proofs, etc.

Exhaustive upward generation of KS sets
\cite{pmmm05a,pmmm05a-corr} faces computational
limits of supercomputers and is limited to ca.~40 hypergraph
vertices in 3D, to ca.~25 ones in 4D, and so on.  For these there
is the additional task of rejecting hypergraphs that do not admit
a vector coordinatization, a problem whose existing algorithms are
still computationally infeasible in some cases. However, this
approach remains the only deterministic and completely exhaustive
generation method for obtaining KS sets. 

Polytopes or Pauli operators can serve as providers of big master
sets which enable random downward generation of smaller sets
\cite{aravind10,waeg-aravind-jpa-11,mfwap-11,mp-nm-pka-mw-11,waeg-aravind-megill-pavicic-11,waegell-aravind-12,waeg-aravind-pra-13,waeg-aravind-fp-14,waeg-aravind-jpa-15,waeg-aravind-pla-17,pavicic-pra-17}.
However, this method is based on serendipitous or intuitively found
polytopes or operators. Also, they do not generalize. Still, they
are a valuable source of coordinatization and many KS set features
as we also show below \cite{pavicic-pra-17}.

In the present paper, we present a new method of generating KS
sets from a small set of components of orthogonal vectors.
In other words, we show that sets of simple orthogonal vectors
inherently lead to KS sets. Such simple components we obtain
either from the coordinatization of the master sets from the
aforementioned polytope approach or directly from an automated
computer search. It provides us not only with a uniform and
general method for KS set generation but also with a larger scope
and a bigger, more thorough picture of quantum contextuality than
any of the previous approaches.

Most of the results in the paper are generated within our hypergraph
language and its algorithms and programs written
in C we developed in
\cite{bdm-ndm-mp-1,pmmm05a-corr,pm-ql-l-hql2,pmm-2-10,bdm-ndm-mp-fresl-jmp-10,mfwap-11,mp-nm-pka-mw-11,megill-pavicic-mipro-17s}
and extended here, as well as the parity-proof algorithms and
programs developed in
\cite{aravind10,waeg-aravind-jpa-11,waeg-aravind-megill-pavicic-11,waeg-aravind-jpa-15}.

In the end, we discuss several new features and interconnectedness
of the KS sets from the 4D and 6D KS classes which serve us as
examples of our method.

The paper is organized as follows. Formalism, algorithms, and
programs are introduced in Sec.~\ref{sec:mmp}. A small KS set master
is vector generated in Sec.~\ref{sec:101}, where also a method of 
stripping edges and generating a KS class is introduced.
A general vector generation of large, so far unknown, KS master sets
and KS classes is presented in Sec.~\ref{sec:big}.
Sec.~\ref{sec:control} deals with controlling the sizes of vector
generated sets. In Sec.~\ref{sec:terr} we generated a new 6D class
which contains the so-called {\em star\/} set, herewith closing a
related open question. Sec.~\ref{sec:con} contains our conclusions.

\section{Contextual hypergraph language---algorithms
  and programs}
\label{sec:mmp}

We encode KS sets via McKay-Megill-Pavi\v ci\'c (MMP)
hypergraphs which are defined as hypergraphs in which edges that
intersect each other in $n-2$ vertices contain at least $n$ vertices.
Vertices correspond to vectors and edges to their orthogonalities.
The MMP hypergraph formalism has been developed in
\cite{pmmm05a-corr,pavicic-pra-17,megill-pavicic-mipro-17s}.

The essence of the KS formalism lies in the inability to assign
predetermined values, 1 and 0, to vertices and edges
(vectors and their orthogonal 4-tuples) of MMP hypergraphs,
graphically visualised as dots and lines (straight or curved)
connecting each (orthogonal) four of them. They form sets, called
KS sets, in which the following condition is violated: ``One and
only one of the vertices from any edge of the set is assigned 1,
while the others are assigned 0.'' Each KS set is thus represented
by a collection of edges mutually connected into a single hypergraph
which, by its very design, amounts to a proof of the Kochen-Specker
theorem, provided its coordinatization can be given. 
\cite{koch-speck,pavicic-pra-17}.

MMP hypergraphs are encoded by means of printable ASCII characters.
Vertices are denoted by one of the
following characters: {{\tt 1 2 \dots 9 A B \dots Z a b
\dots z ! " \#} {\$} \% \& ' ( ) * - / : ; \textless\ =
\textgreater\ ? @ [ {$\backslash$} ] \^{} \_ {`} {\{}
{\textbar} \} $\sim$ \cite{pmm-2-10}. When all of
them are exhausted one reuses them prefixed by `+',
then again by `++', and so forth. A number of examples are given
below. We generate, process, and handle MMP hypergraphs by means
of algorithms in the programs
SHORTD, MMPSTRIP, MMPSUBGRAPH, VECFIND, STATES01, and others
\cite{bdm-ndm-mp-1,pmmm05a-corr,pmm-2-10,bdm-ndm-mp-fresl-jmp-10,mfwap-s-11,mp-nm-pka-mw-11}. We have also defined extensions to the notation (prefixes
and suffixes) that allow us to attach additional information such
as a vector assignment to the vertices \cite{megill-pavicic-mipro-17s}.
All of our programs work seamlessly within this {\em language} that
unambiguously describes hypergraphs and associated information.
Our programs are available for
general use in our repository \cite{master-repository-17}.

\section{\{-1,0,1\}-component generation of the
  smallest KS class}
\label{sec:101}

Our program VECFIND provides us with master sets generated by
vector components. As shown in \cite{mp-mw-nm-pka-18}
\{-1,0,1\} components give us a master set with 40 vertices and
32 edges, denoted as 40-32. It consists of two disconnected
subsets: a 24-24 KS set and a 16-8 non-KS set, as shown in
Fig.~\ref{fig:ker}. The 24-24 set is isomorphic with the 24-24
set found by A.~Peres in 1991 \cite{peres}. Had Peres recognized 
that his set can be given a graphical representation as in
Fig.~\ref{fig:ker}, he would have immediately seen all the smaller
sets contained in it. Without it, it took three years until
M.~Kernaghan \cite{kern} found one of the subsets.

\begin{widetext}
\begin{figure}[ht]
\centering
\includegraphics[width=0.99\textwidth]{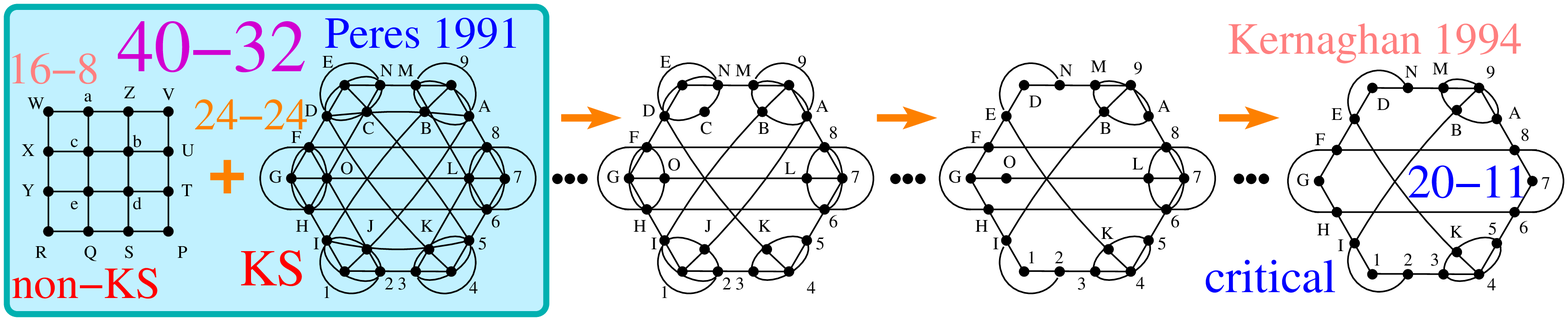}
\caption{40-32 master set obtained from \{-1,0,1\} vector components
  via our program VECFIND. Stripping of edges from the 24-24
  KS set it contains gives us Kernaghan's 20-11 in a few steps.}
\label{fig:ker}      
\end{figure}
\end{widetext}

Two years later, A.~Cabello et al.~\cite{cabell-est-96a} found
another KS set (18-9) also from Peres' 24-24. Note that the two
24-24s from Figs.~\ref{fig:ker} and \ref{fig:cab} are isomorphic
to each other.

\begin{widetext}
\begin{figure}[ht]
\centering
\includegraphics[width=0.99\textwidth]{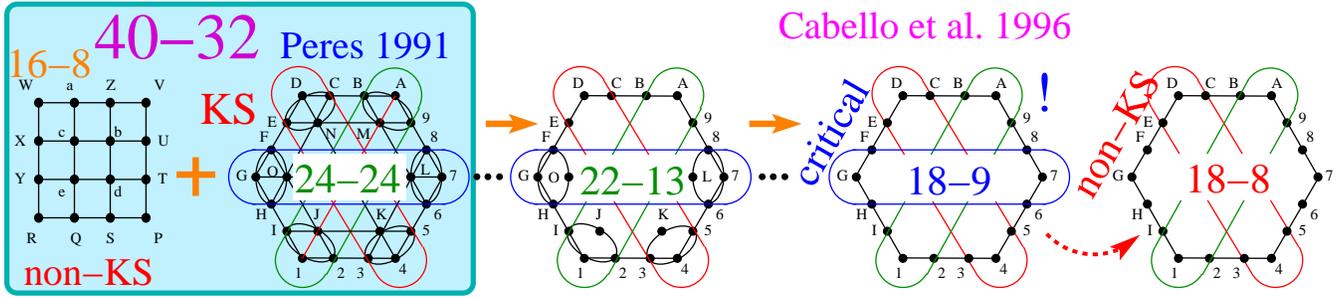}
\caption{Another isomorphic representation of the 24-24 KS set
  contained in 40-32 master set. Stripping of its edges gives us
  Cabello et al.'s 18-9 in a few steps.}
\label{fig:cab}      
\end{figure}
\end{widetext}

In 2005 we carried out an exhaustive constructive upward
hypergraph-generation of KS sets of up to 23 edges and a single
24-24, i.e., Peres' set \cite{pmmm05a}. In that paper we extracted
all KS sets whose vertices could be given a coordinatization with
\{-1,0,1\} components. We also proved that the aforementioned 18-9
is the smallest 4D KS set with such a coordinatization.  

Simple vector components, in our case \{-1,0,1\},
introduced as inputs into our program exhaustively gives a
KS master set 40-32. Exhaustively here means that all vectors
with the given components are used up for the construction of
the master. That also means that the 24-24 set does not make use
of all vectors with these components. Some serve to build the
non-KS 16-8 also contained in the 40-32.

Graphical representation of MMP hypergraphs first given in
\cite{pmmm05a} and shown in Figs.~\ref{fig:ker} and \ref{fig:cab} 
prompted us to design a stripping algorithm and program MMPSTRIP
which served us to strip the edges from the 24-24 KS set and obtain
all 1232 KS sets contained in it within less than 2 mins on a PC
\cite{pmm-2-10}. We say that such KS subsets of a master KS set
form a KS {\em class}. 24-24 contains many more non-KS subsets
and we filter out the KS subsets by means of our program STATES01.

Thus our method consists of, first, generating a KS master set
from simple vector components by means of VECFIND, then stripping
its edges with MMPSTRIP, and finally filtering out its KS subsets 
via STATES01 so as to keep only minimal KS subsets; minimal, in the
sense that a removal of any edge (i.e., any $n$-tuple of mutual
orthogonalities, of $n$ vectors from an $n$-dimensional Hilbert
space), turns such a KS subset into a non-KS set. In other words,
they represent a KS setup that has no redundancy. We call these
{\em critical\/} KS sets. They are all we need for an
experimental implementation---additional orthogonalities that 
bigger KS sets that contain critical ones might possess do not add
any new property to the ones that the minimal critical core already
has.

\section{Generation of bigger KS classes}
\label{sec:big}

Bigger KS classes were so far mostly generated from the master
sets obtained with the help of polytopes
\cite{aravind10,waeg-aravind-jpa-11,mfwap-11,mp-nm-pka-mw-11,waeg-aravind-megill-pavicic-11,waegell-aravind-12,waeg-aravind-pra-13,waeg-aravind-fp-14,waeg-aravind-jpa-15,waeg-aravind-pla-17,pavicic-pra-17}.
In this section we shall consider an extension of the 148-265 KS
class we obtained in \cite{pavicic-pra-17} from the 148-265 master
KS set which Waegell and Aravind obtained from the Witting polytope
\cite{waeg-aravind-pla-17}.

They made use of the vector components
$\{0,\pm i,\pm 1,\pm\omega,\break \pm\omega^2,\pm i\omega^{1/\sqrt{3}},\pm i\omega^{2/\sqrt{3}}\}$ (where $\omega$ is the cubic root of unity,
$\omega=e^{2\pi i/3}$), which they derived from the Witting polytope.
In \cite{pavicic-pra-17} we showed that the 
set of components $\{0,\pm 1,\pm\omega,\pm\omega^2\}$
suffices for the coordinatization of the 148-265.

But even this smaller set of components immediately reveals
that the master set we would obtain by our vector generation
method must contain the 40-32 class from Sec.~\ref{sec:101}
while the polytope-derived 148-265 class does not contain KS sets
smaller than the 40-23 critical KS set, as one can see from Fig.~10
in \cite{pavicic-pra-17}. Thus the master set turns out to be
much larger than the polytope-generated one---it has 400
vertices and 1012 edges and that 400-1012 master set does
not split into disconnected subsets as the 40-32 master set. By
means of MMPSTRIP and STATES01 we then extract smaller KS subsets
contained in the master set. A distribution of the smallest
critical subsets is shown in Fig.~\ref{fig:om-distr},
and we see that it populates the span between the 40-32 class
whose maximal critical is 24-15 and the polytope 148-265 class
whose minimal critical is 40-23.

\begin{figure*}[ht]
\centering
\includegraphics[width=0.7\textwidth,height=0.4\textwidth]{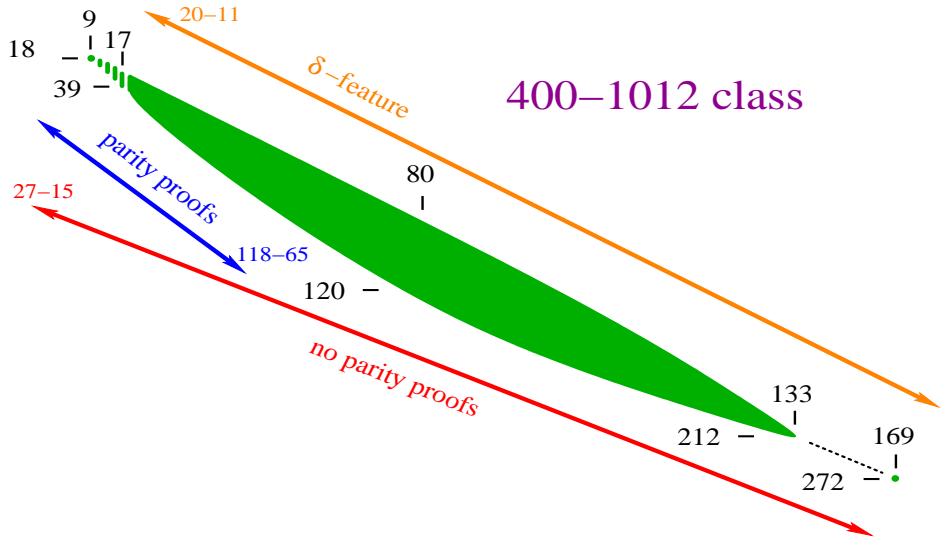}
\caption{Distribution of $38,132$ critical KS sets from the
  400-1012 class we found by means of programs MMPSTRIP and
  STATES01}
\label{fig:om-distr}      
\end{figure*}

\begin{figure*}[ht]
\centering
\includegraphics[width=0.99\textwidth]{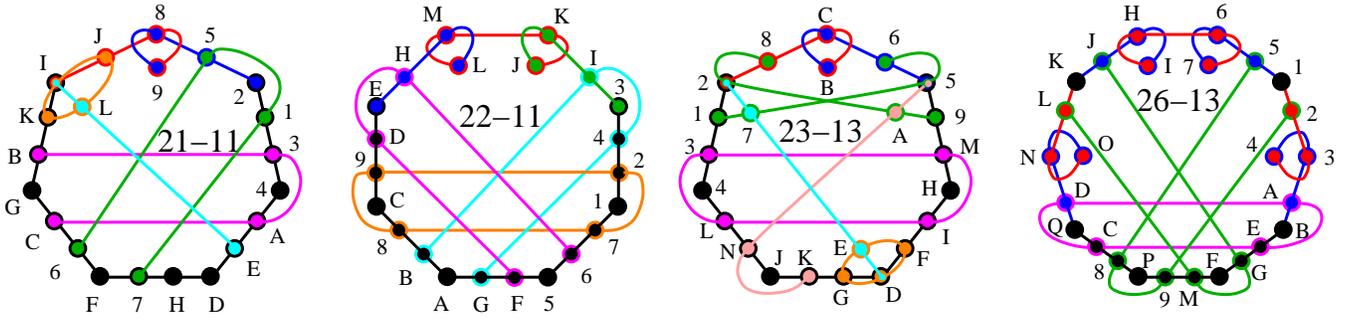}
\caption{Some of the smallest KS MMP hypergraphs from
  400-1012 class that cannot be generated from \{-1,0,1\}.}
\label{fig:om-figs}      
\end{figure*}

Only smaller criticals with up to 118 vertices and 65 edges
have parity proofs while most of the sets starting with 27-15
lack it. Also practically all KS sets apart from the smallest
one 18-9 possess the so-called $\delta$-feature of two edges
sharing two vertices, i.e., intersecting each other at two
vertices \cite{pavicic-pra-17}. This is in contrast to the
300-675 class \cite{pavicic-pra-17} where none of $10^{10}$
KS criticals possesses the $\delta$-feature.

In Fig.~\ref{fig:om-figs} we present MMP hypergraphs of some
of the smallest critical KS sets from the 400-1012 class. The
first three are of the same size as the sets from 40-32 class but
they cannot be generated from the \{-1,0,1\} vector components.
Their MMP hypergraph strings and coordinatizations read as follows.

{\setstretch{0.8}{
{\bf 21-11} {\small 1234,1567,2589,A3BC,ADE4,F6GC,F7DH,IJ89,IJKL,\break IKGB,ILEH.\{1=\{0,0,0,1\},2=\{0,0,1,0\},A=\{0,0,1,1\},5=\{0,1,0,0\},F=\{0,\break 1,0,$\omega$\},I=\{0,1,$\omega$,0\},J=\{0,1,-$\omega$,0\},8=\{1,0,0,1\},9=\{1,0,0,-1\},K=\{$\omega$,0,0,\break 1\},L=\{$\omega$,0,0,-1\},6=\{$\omega$,0,1,0\},7=\{$\omega$,0,-1,0\},3=\{1,$\omega$,0,0\},D=\{$\omega$,$\omega^2$,1,-1\},E=\{$\omega$,$\omega^2$,-1,1\},G=\{$\omega$,$\omega^2$,-1,-1\},4=\{1,-$\omega$,0,0\},H=\{$\omega$,-$omega^2$,1,1\},\break B=\{$\omega$,-$\omega^2$,1,-1\},C=\{$\omega$,-$\omega^2$,-1,1\}\}}

{\bf 22-11} {\small 1234,1567,8927,8ABC,9DEC,5FAG,6FDH,I3JK,I4BG,\break EHLM,LJKM.\{1=\{0,0,1,0\},8=\{0,0,1,$\omega$\},9=\{0,0,1,-$\omega$\},2=\{0,1,0,0\},5=\break \{0,$\omega$,0,1\},6=\{0,$\omega$,0,-1\},I=\{0,1,$\omega$,0\},7=\{1,0,0,0\},3=\{1,0,0,1\},4=\{1,0,0,\break-1\},F=\{$\omega$,0,-1,0\},D=\{1,$\omega$,$\omega^2$,1\},A=\{1,$\omega$,$\omega^2$,-1\},B=\{1,$\omega$,-$\omega^2$,1\},E=\{1,\break $\omega$,-$\omega^2$,-1\},C=\{1,-$\omega$,0,0\},G=\{1,-$\omega$,$\omega^2$,1\},H=\{1,-$\omega$,$\omega^2$,-1\},L=\{1,$\omega^2$,1,\break 1\},J=\{1,$\omega^2$,-1,-1\},K=\{1,-$\omega^2$,1,-1\},M=\{1,-$\omega^2$,-1,1\}\}}

{\bf 23-13} {\small 1234,1567,829A,82BC,DE27,DEFG,DFHI,DJKG,L3MI,\break L4JN,56BC,59MH,5AKN.\{1=\{0,0,0,1\},8=\{0,0,1,0\},D=\{0,0,$\omega$,1\},E=\break \{0,0,$\omega$,-1\},2=\{0,1,0,0\},L=\{0,1,0,1\},5=\{0,1,$\omega$,0\},6=\{0,1,$-\omega$,0\},7=\{1,\break 0,0,0\},9=\{1,0,0,1\},A=\{1,0,0,-1\},B=\{$\omega$,0,0,1\},C=\{$\omega$,0,0,$-1$\},3=\{1,0,\break $\omega$,0\},4=\{1,0,$-\omega$,0\},F=\{1,1,0,0\},J=\{1,1,$\omega$,-1\},K=\{1,1,-$\omega$,1\},M=\{1,1,\break -$\omega$,-1\},G=\{1,-1,0,0\},N=\{1,-1,$\omega$,1\},H=\{1,-1,$\omega$,-1\},I=\{1,-1,-$\omega$,1\}\}}

{\bf 26-13} {\small 1234,1567,2589,AB34,ACDE,BFGE,HIJK,HI67,LJGM,\break LKNO,P8CQ,P9FM,NODQ.\{1=\{0,0,0,1\},2=\{0,0,1,0\},A=\{0,0,1,$\omega$\},\break B=\{0,0,1,-$\omega$\},5=\{0,1,0,0\},H=\{0,1,0,$\omega$\},I=\{0,1,0,-$\omega$\},L=\{0,$\omega$,0,1\},P=\break \{0,1,-$\omega$,0\},8=\{1,0,0,$\omega$\},9=\{1,0,0,-$\omega$\},J=\{1,0,1,0\},K=\{1,0,-1,0\},6=\break \{$\omega$,0,1,0\},7=\{$\omega$,0,-1,0\},N=\{1,1,1,-$\omega^2$\},O=\{1,-1,1,$\omega^2$\},3=\{1,$\omega$,0,0\},\break 4=\{1,-$\omega$,0,0\},F=\{1,$\omega^2$,1,$\omega$\},C=\{1,$\omega^2$,1,-$\omega$\},D=\{1,$\omega^2$,-1,$\omega$\},G=\{1,$\omega^2$,\break -1,-$\omega$\},E=\{$\omega$,-1,0,0\},M=\{1,-$\omega^2$,-1,$\omega$\},Q=\{-1,$\omega^2$,1,$\omega$\}\}}

\section{Controlling the size of  KS classes}
\label{sec:control}
}}

In Sec.~\ref{sec:big} we obtained a big KS class from the
set of components $\{0,\pm 1,\pm\omega,\pm\omega^2\}$ which closed
the gap from the 40-32 (24-24) and 148-256 KS classes. However,
if we just wanted to generate smaller KS sets that would include
sets from that gap, then starting with a big master set is a waste
of resources and slows down the generation.

To resolve such a problem we consider the components and choose
the most efficient ones either by an automated selection or by
examination of components. For instance, we see that \{-1,0,1\}
and ($-\omega$,0,$\omega$) when considered alone are equivalent. 
Combined, they form a bigger set but if we dispense with \{-1,0,1\}
and just combine $\pm\omega$ and $+\omega^2$, so as to make use of
\{$-\omega$,0,$\omega$,$\omega^2$\} components, we obtain a fairly
small 180-203 KS master set which can be processed within hours
on a supercomputer. A distribution of its criticals 
is given in Fig.~\ref{fig:oms-distr}. 

\begin{figure}[h]
\centering
\includegraphics[width=0.49\textwidth]{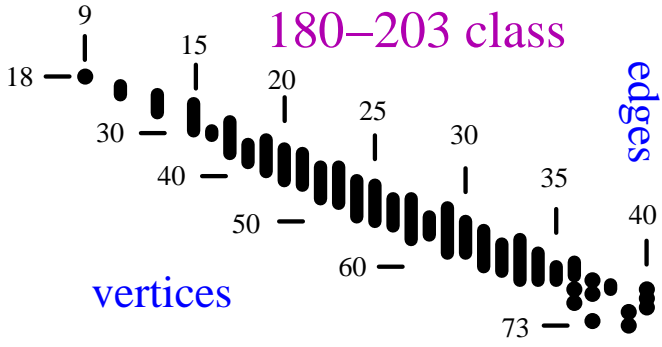}
\caption{Distribution of $96,173$ critical KS sets from the
  180-202 class. Note the KS critical sets with even number of
  edges, 16 and 18.}
\label{fig:oms-distr}      
\end{figure}

We can see that there are KS criticals with 16 and 18 edges which,
as can be seen from the Fig.~\ref{fig:om-distr}, we still have not
obtained within 400-1012 distribution although we run the generation
of the latter class ten times longer than of the former. 

\bigskip\bigskip

\section{Generation in a new territory}
\label{sec:terr}

The 4D KS space was extensively examined, and many ways of
finding KS sets were explored. Not so in the other dimensions,
though. For example, the first KS set in the 6D was found only
recently \cite{lisonek-14}. It is the star set shown in 
Fig.~\ref{fig:star}(a).

The question emerged on whether we could find a KS 
class that contains the star set. In \cite{pavicic-pra-17},
another huge 6D class was generated, but it did not contain the
star set. It was only shown that there exists a triangular
representation of the star set shown in Fig.~\ref{fig:star}(b).

\begin{widetext}
\begin{figure}[ht]
\centering
\includegraphics[width=0.98\textwidth]{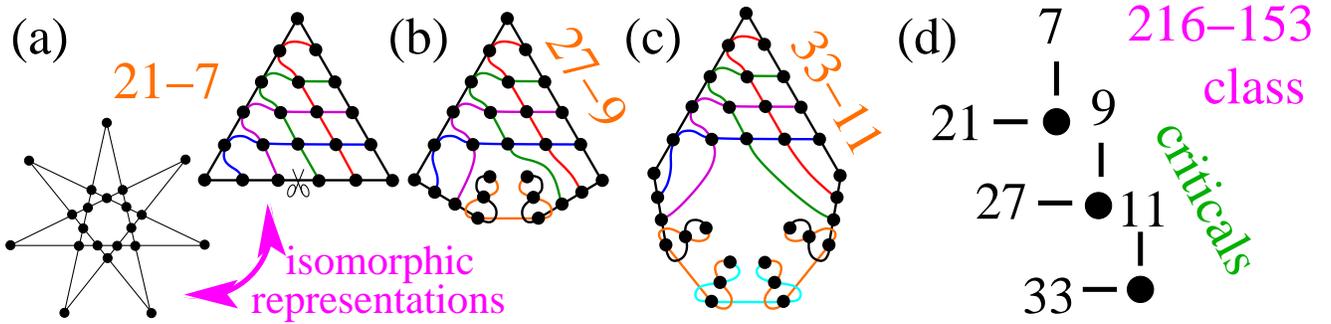}
\caption{Criticals (a,b,c) and a distribution (d) of the 216-153 KS
  class; these criticals are the only critical KS sets contained
  in the 216-153 master set.}
\label{fig:star}      
\end{figure}
\end{widetext}

In \cite{lisonek-14}, the coordinatization of the star set was
based on \{0,1,$\omega$,$\omega^2$\} components. In
\cite{pavicic-pra-17} it was shown that \{0,1,$\omega$\} suffice.
But only the present vector generation enables us to generate
the master set 216-153 from these components. It contains
a huge number of KS subsets, but surprisingly only three critical
sets 21-7 (star/triangle set), 27-9, and 33-11 shown as
(a), (b), and (c) in Fig.~\ref{fig:star}, respectively. 

\section{Conclusions}
\label{sec:con}

This work presents an algorithm to find Kochen-Specker configurations
that admit a coordinatization with a pre-chosen set of vector
components and dimension. The algorithm allows researchers to design
KS configurations that match their desired experimental setup, rather
than being constrained to shoehorn their experiments into a handful
of serendipitous configurations available in the literature. This
gives the experimenter unprecedented freedom to exploit contextuality
in an optimal way for use in quantum computation and communication.

\section*{Acknowledgments}
\label{sec:ack}

Supported by the Croatian Science Foundation 
project IP-2014-09-7515, the Ministry of Science and Education 
of Croatia through the Center of Excellence CEMS funding, 
grants Nos. KK.01.1.1.01.0001 and 533-19-15-0022 and the Alexander
von Humboldt Foundation.
Computational support was provided by the cluster Isabella of
the Zagreb University Computing Centre, by the Croatian National
Grid Infrastructure (CRO-NGI), and by the Center for Advanced
Computing and Modelling (CNRM) for providing computing resources
of the supercomputer Bura at the University of Rijeka in Rijeka,
Croatia. The supercomputer Bura and other information and 
communication technology research infrastructure were acquired
through the project
{\it Development of research infrastructure for laboratories of
the University of Rijeka Campus}, which is co-funded by the
European regional development fund.

\end{document}